\documentclass[twocolumn]{aastex63}

\usepackage{color}

\usepackage{enumitem}
\usepackage{amsmath}

\newcommand{\be}{\begin{equation}}
\newcommand{\ee}{\end{equation}}

\newcommand{\wmap}{\textsl{WMAP}}

\newcommand{\planck}{\textsl{Planck}}
\newcommand{\Planck}{\textsl{Planck}}
\newcommand{\lcdm}{\ensuremath{\Lambda\mathrm{CDM}}}

\providecommand{\sorthelp}[1]{}

\begin{document}

\title{High $H_0$ Values from CMB E-mode Data: A Clue for Resolving the Hubble Tension?}

\correspondingauthor{Graeme E. Addison}
\email{gaddison@jhu.edu}

\author[0000-0002-2147-2248]{Graeme~E.~Addison}
\affil{Dept. of Physics \& Astronomy, The Johns Hopkins University, 3400 N. Charles St., Baltimore, MD 21218-2686}

\shorttitle{High $H_0$ from E-mode data}
\shortauthors{G.~E.~Addison} 

\begin{abstract}

\noindent
The E-mode (EE) CMB power spectra measured by \planck, ACTPol, and SPTpol constrain the Hubble constant to be $70.0\pm2.7$, $72.4^{+3.9}_{-4.8}$, and $73.1^{+3.3}_{-3.9}$~km~s$^{-1}$~Mpc$^{-1}$ within the standard \lcdm\ model (posterior mean and central 68\% interval bounds). These values are higher than the constraints from the \planck\ temperature (TT) power spectrum, and consistent with the Cepheid-supernova distance ladder measurement $H_0=73.2\pm1.3$~km~s$^{-1}$~Mpc$^{-1}$. If this preference for a higher value was strengthened in a joint analysis it could provide an intriguing hint at the resolution of the Hubble disagreement. We show, however, that combining the \planck, ACTPol, and SPTpol EE likelihoods yields $H_0=68.7\pm1.3$~km~s$^{-1}$~Mpc$^{-1}$, $2.4\sigma$ lower than the distance ladder measurement. This is due to different degeneracy directions across the full parameter space, particularly involving the baryon density, $\Omega_bh^2$, and scalar tilt, $n_s$, arising from sensitivity to different multipole ranges. We show that the E-mode \lcdm\ constraints are consistent across the different experiments within $1.4\sigma$, and with the \planck\ TT results at $0.8\sigma$. Combining the \planck, ACTPol, and SPTpol EE data constrains the phenomenological lensing amplitude, $A_L=0.89\pm0.10$, consistent with the expected value of unity.

\end{abstract}

\keywords{\href{http://astrothesaurus.org/uat/322}{Cosmic microwave background radiation (322)}; \href{http://astrothesaurus.org/uat/758}{Hubble constant (758)}; \href{http://astrothesaurus.org/uat/1146}{Observational cosmology (1146)}}

\section{Introduction}
\label{sec:intro}

The standard Lambda-Cold Dark Matter (\lcdm) cosmological model is supported by a range of observations, including fluctuations in the cosmic microwave background (CMB), the large-scale distribution of galaxies, and, with the exception of lithium-7, primordial element abundance \citep[e.g.,][]{bennett/etal:2013,planck2016-l06,fields/etal:2020,alam/etal:prep}. A significant disagreement has emerged between \lcdm\ constraints and more direct measurements of the Hubble constant, $H_0$. Combining \planck\ CMB data with baryon acoustic oscillation (BAO) measurements yields $H_0=67.61\pm0.44$~km~s$^{-1}$~Mpc$^{-1}$ \citep{alam/etal:prep}, while the latest Cepheid-supernova distance ladder result is $73.2\pm1.3$~km~s$^{-1}$~Mpc$^{-1}$ from the Supernova $H_0$ for the Equation Of State team \citep[SH0ES;][]{riess/etal:2021}. The SH0ES result is supported by various partially or fully independent low-redshift observations \citep[e.g.,][]{wong/etal:2020,schombert/mcgaugh/lelli:2020,ehsan/etal:2020,soltis/casertano/riess:2021,blakeslee/etal:prep}. Some low-redshift analyses have yielded lower values, consistent with both the CMB+BAO value and the SH0ES result \citep[e.g.,][]{freedman/etal:2019,birrer/etal:2020}. The `low' values of $H_0$ based on measurements sensitive to the physics in the early universe are not driven by any single data set or observational method \citep[e.g.,][]{aubourg/etal:2015,addison/etal:2018}.

CMB data play a critical role in constraining possible deviations from \lcdm, including modifications introduced in an attempt to address the $H_0$ disagreement \citep[e.g.,][and references therein]{knox/millea:2020}. Future improvements in cosmological constraints from the CMB will largely come from polarization. The \planck\ temperature-polarization (TE) cross-spectrum already constrains \lcdm\ parameters with a comparable precision to the temperature (TT) spectrum, for example $H_0=68.44\pm0.91$~km~s$^{-1}$~Mpc$^{-1}$ from TE+\texttt{lowE}, and $66.88\pm0.92$~km~s$^{-1}$~Mpc$^{-1}$ from TT+\texttt{lowE}, where `\texttt{lowE}' denotes the E-mode polarization likelihood at $\ell<30$, which primarily constrains the optical depth, $\tau$ \citep{planck2016-l06}.

Ongoing and upcoming ground-based surveys including Advanced ACTPol \citep{henderson/etal:2016}, SPT-3G \citep{benson/etal:2014}, Simons Observatory \citep{simonsobservatory:2019}, and CMB-S4 \citep{stage4:2019}, will make increasingly precise measurements of the EE spectrum, ultimately aiming to achieve uncertainties dominated by signal (sample variance) in the damping tail out to $\ell\simeq3000-5000$. By this point, the E-mode polarization will be more constraining than the temperature fluctuations, both for \lcdm\ parameters like $H_0$, and additional parameters constrained from the damping tail, such as the effective number of relativistic species, $N_{\rm eff}$ (e.g., \citealt{galli/etal:2014}, Section~4 of \citealt{simonsobservatory:2019}).

Parameter uncertainties from current EE spectra from \planck, the Atacama Cosmology Telescope (ACT) ACTPol receiver \citep{thornton/etal:2016}, and the South Pole Telescope (SPT) SPTpol camera \citep{austermann/etal:2012} are fairly large, with $H_0$ constrained to 4-6\% precision for each survey \citep{planck2016-l06,henning/etal:2018,aiola/etal:2020}. Recently, \cite{dutcher/etal:prep} reported the first cosmological constraints from the SPT-3G receiver. Assuming \lcdm, they found a preference for a higher $H_0$ from the EE spectrum ($76.4\pm4.1$~km~s$^{-1}$~Mpc$^{-1}$) than \planck\ TT. They pointed out that the EE results from every one of \planck, ACTPol, SPTpol, and SPT-3G individually exhibit this same trend (see their Fig.~13). The EE spectra mildly prefer higher $H_0$ values than the \planck\ TT data, and also their TE counterparts.

Is this trend made more significant by combining the EE spectra from the different surveys? If so, it could be hinting at some modification to \lcdm\ that addresses the Hubble tension while impacting temperature and polarization results differently. Such a hint would be valuable since the low-redshift $H_0$ measurements provide very little direction for physical resolutions of the disagreement. Alternatively, statistically significant shifts in parameters from EE compared to TT could indicate some new systematic issue that needs to be understood moving forward.

In this work we reexamine constraints from the \planck, ACTPol and SPTpol EE data\footnote{At the time of writing, the SPT-3G likelihood from \cite{dutcher/etal:prep} is not publicly available, although we expect the SPT-3G  and SPTpol results to produce similar constraints when combined with \planck\ (see Section~\ref{sec:resultscombined}).} to shed light on this issue, motivated by both parameter tensions and the importance of the EE measurements for the future of CMB cosmology. We describe the public data sets and codes used in this work in Section~\ref{sec:datamodel}, present results in Section~\ref{sec:results} and conclude in Section~\ref{sec:conclusions}. The main results are shown in Table~\ref{table:results} and Figure~\ref{fig:results}.

\section{Data and Model Fitting}
\label{sec:datamodel}

We perform cosmological parameter fitting using Markov chain Monte Carlo (MCMC) sampling implemented in the \texttt{CosmoMC}\footnote{https://cosmologist.info/cosmomc/} package \citep{lewis/bridle:2002,lewis:2013}. Theoretical CMB power spectra are computed from parameters using \texttt{CAMB}\footnote{https://camb.info/} \citep{lewis/challinor/lasenby:2000,howlett/etal:2012}. The \lcdm\ parameters varied in the fits are the physical baryon and cold dark matter densities, $\Omega_bh^2$ and $\Omega_ch^2$, the \texttt{CosmoMC} parameter $\theta_{\rm MC}$, which is closely related to the angular spacing of the acoustic peaks, the scalar amplitude, $A_s$, the optical depth, $\tau$, and the scalar tilt, $n_s$. The value of $H_0$ at each step in the chain is then derived from these. We follow the assumptions used by \cite{planck2016-l06} for massive neutrinos (fixing sum of masses to $0.06$~eV), the connection between primordial helium abundance and $\Omega_bh^2$, and the parametric `tanh' form for the evolution of the ionization fraction at reionization.

The \planck\ and ACTPol teams have released multiple likelihood versions. In this work we use the likelihoods used to compute the main results reported by these collaborations. For \planck\ we use the \texttt{plik} likelihood at $\ell\geq30$, and the \texttt{lowE} likelihood at $\ell<30$. These likelihoods are described by \cite{planck2016-l05} and available on the Planck Legacy Archive\footnote{http://pla.esac.esa.int/pla/}. For ACTPol we use the public \texttt{ACTPollite} likelihood described by \cite{choi/etal:2020} and \cite{aiola/etal:2020}\footnote{https://lambda.gsfc.nasa.gov/product/act/act\_dr4\_likelihood\_get.cfm}. For SPTpol we use the public likelihood provided by \cite{henning/etal:2018}\footnote{https://pole.uchicago.edu/public/data/henning17/ and https://lambda.gsfc.nasa.gov/product/spt/sptpol\_lh\_2017\_get.cfm}. Foreground and nuisance parameters are varied in the fits using priors recommended by each collaboration and marginalized over in all results provided in this work. The overall calibration of the \planck, ACTPol, and SPTpol spectra, and associated uncertainties, are discussed in Section~3.3.4 of \cite{planck2016-l05}, Section~7 of \cite{choi/etal:2020}, and Section~4.5 of \cite{henning/etal:2018}, respectively. Convergence of the fits is assessed using multiple chains following the standard \texttt{CosmoMC} approach, requiring the Gelman-Rubin diagnostic of within-chain and across-chain parameter spread, $R-1<0.01$ \citep{gelman/rubin:1992}.

\section{Results}
\label{sec:results}

\begin{table*}
  \centering
  \caption{Posterior mean and central 68\% \lcdm\ parameter constraints from fitting to EE spectra, including in each case the $\ell<30$ \planck\ \texttt{lowE} likelihood to constrain $\tau$, with $H_0$ reported in km~s$^{-1}$~Mpc$^{-1}$}
  \begin{tabular}{lcccccc}
\hline
\hline
EE Data&$100\Omega_bh^2$&$\Omega_ch^2$&$100\theta_{\rm MC}$&$10^9A_se^{-2\tau}$&$n_s$&$H_0$\\
\hline
\planck & $2.40\pm0.12$ &  $0.1157^{+0.0043}_{-0.0047}$ & $1.04000^{+0.00087}_{-0.00086}$ & $1.905\pm0.024$ &  $0.980^{+0.013}_{-0.015}$ & $70.0\pm2.7$ \\
ACTPol & $2.27^{+0.12}_{-0.14}$ & $0.108\pm0.011$ &  $1.0409^{+0.0015}_{-0.0016}$ & $1.92^{+0.21}_{-0.20}$ & $0.986^{+0.053}_{-0.060}$ & $72.4^{+3.9}_{-4.8}$\\
SPTpol & $2.25\pm0.12$ &  $0.1055^{+0.0077}_{-0.0076}$ & $1.0408\pm0.0016$ &  $1.722\pm0.068$ &  $1.029^{+0.038}_{-0.048}$ & $73.1^{+3.3}_{-3.9}$\\
\hline
\Planck+ACTPol & $2.274^{+0.061}_{-0.062}$ & $0.1187\pm0.0032$ &  $1.03988\pm0.00068$ &  $1.884\pm0.022$ &  $0.9663^{+0.0100}_{-0.0101}$ & $67.8\pm1.6$ \\
\Planck+SPTpol & $2.314\pm0.060$ &  $0.1165\pm0.0030$ &  $1.03997\pm0.00075$ &  $1.877\pm0.021$ &  $0.975\pm0.010$ &  $68.9\pm1.5$ \\
ACTPol+SPTpol & $2.268^{+0.082}_{-0.083}$ & $0.1070^{+0.0057}_{-0.0056}$ & $1.04087^{+0.00101}_{-0.00099}$ & $1.762^{+0.059}_{-0.060}$ & $0.998^{+0.028}_{-0.032}$ & $72.6^{+2.3}_{-2.6}$\\
\hline
\Planck+ACTPol+SPTpol & $2.287\pm0.048$ &  $0.1167\pm0.0027$ &  $1.04002^{+0.00061}_{-0.00060}$ & $1.869\pm0.020$ &  $0.9700\pm0.0094$ &  $68.7\pm1.3$ \\
\hline
\label{table:results}
\end{tabular}
\end{table*}

\subsection{Results from each experiment separately}
\label{sec:resultsseparate}

We present results from fitting \lcdm\ to the \planck, SPTpol, and ACTPol EE spectra in Table~\ref{table:results}. We provide the posterior mean plus bounds of the central 68\% interval. Since the constraint on $\tau$ is largely driven by the \planck\ $\ell<30$ \texttt{lowE} likelihood in each case we report the combination $A_se^{-2\tau}$ rather than $A_s$ and $\tau$ separately \citep{kosowsky/milosavljevic/jiminez:2002}. Figure~\ref{fig:results} shows two-dimensional contours containing 68 and 95\% of the posterior distributions for each experiment. In the figure we show $H_0$ instead of $\theta_{\rm MC}$ to facilitate comparison with low-redshift measurements.

We remark here that, for all three experiments separately, $n_s$ lies within $1.5\sigma$ of unity, and $H_0$ differs by at most $1.1\sigma$ from the latest SH0ES measurement of $73.2\pm1.3$~km~s$^{-1}$~Mpc$^{-1}$ \citep{riess/etal:2021}.

\subsection{Results from combining experiments}
\label{sec:resultscombined}

\begin{figure*}
\centering
\includegraphics[]{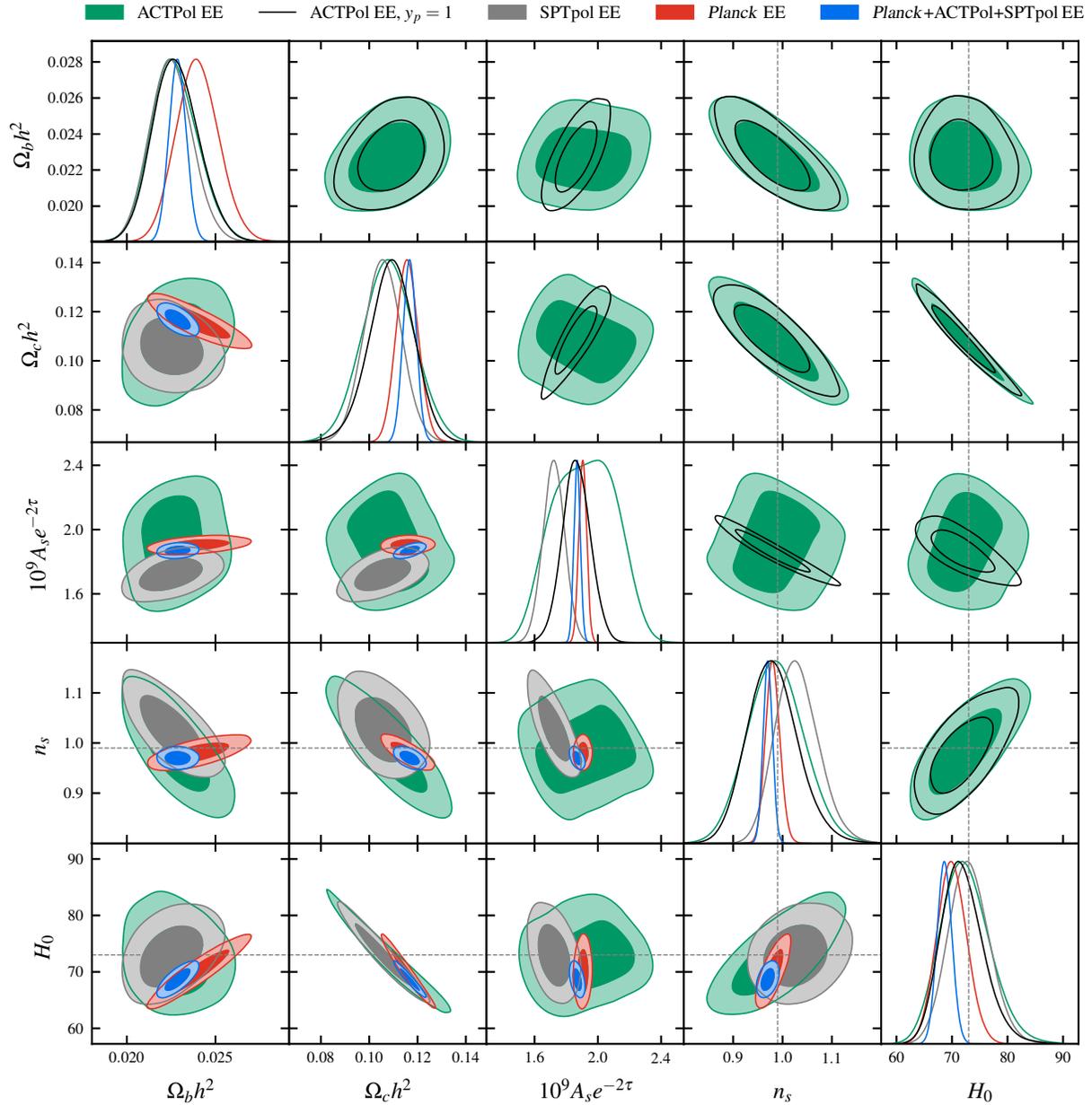}
\caption{Lower triangle: contours containing 68\% and 95\% of the posterior distribution from fits to EE power spectra in \lcdm. Combining the \planck, ACTPol, and SPTpol EE spectra does not reinforce the preference for higher values of $H_0$ from the data sets individually, due to different degeneracy directions, for example in the $n_s-\Omega_bh^2$ plane. Dashed lines at $H_0=73$~km~s$^{-1}$~Mpc$^{-1}$ and $n_s=0.99$ correspond to a point in parameter space that is allowed within the 68\% contours of each experiment individually but disfavored at $3\sigma$ in the combination. Each constraint includes the \planck\ \texttt{lowE} likelihood to constrain $\tau$. Upper triangle: effect of fixing the ACTPol polarization efficiency parameter, $y_p$, as discussed in Section~\ref{sec:resultsconsistencygauss}.}.
\label{fig:results}
\end{figure*}

We show results for \lcdm\ parameters from a joint fit to the \planck, ACTPol, and SPTpol EE spectra in Table~\ref{table:results} and Figure~\ref{fig:results}.

Combining the three EE spectra does not reinforce the preference for higher values of $H_0$ from the individual fits, yielding $H_0=68.7\pm1.3$~km~s$^{-1}$~Mpc$^{-1}$ (posterior mean and central 68\% interval). This value lies $2.4\sigma$ lower than the distance ladder measurement $73.2\pm1.3$~km~s$^{-1}$~Mpc$^{-1}$ \citep{riess/etal:2021}. This is the main result of this paper, and is due to the different degeneracy directions in the full multidimensional \lcdm\ parameter space, and the fact that there are offsets in the preferred mean values from each experiment (even though, as discussed in Section~\ref{sec:resultsconsistency}, below, they are statistically consistent).

To help illustrate this, we include dashed lines in Figure~\ref{fig:results} corresponding to $H_0$ of 73~km~s$^{-1}$~Mpc$^{-1}$ and $n_s$ of 0.99. These values are allowed within the 68\% contours of each individual experiment but disfavored at around $3\sigma$ in the combination. Looking at the $n_s-\Omega_bh^2$ panel of Figure~\ref{fig:results} we see that these parameters are positively correlated for \planck\ EE, but negatively correlated for ACTPol and SPTpol. While the experiments individually prefer or allow higher $n_s$ values, with correspondingly higher $H_0$, this is not possible in the combination because a consensus must also be reached for $\Omega_bh^2$. The portion of multidimensional parameter space most acceptable to all the data sets instead features lower $H_0$ and $n_s$ values. 

Most of the \planck\ EE spectrum constraining power is at larger angular scales ($\ell\lesssim600$), where, for pivot scale $k_p=0.05$~Mpc$^{-1}$, increasing $n_s$ leads to an overall suppression of power. The ACTPol and SPTpol EE constraining power is mostly from finer scales, where increasing $n_s$ leads to an overall enhancement of power. The \planck, ACTPol, and SPTpol EE bandpowers are shown in Figure~8 of \cite{dutcher/etal:prep}. This difference in behavior leads to a significant change in the $n_s-\Omega_bh^2$ degeneracy direction because the derivative of the theory spectrum with respect to $\Omega_bh^2$ also changes sign between larger and smaller angular scales \citep[Fig.~1 of][]{galli/etal:2014}. A similar change in the $n_s-\Omega_bh^2$ degeneracy direction is also apparent over different multipole ranges of the TT spectrum \citep[e.g., Fig.~1 of][]{addison/etal:2016}. These changes in degeneracy directions are reproduced in Fisher forecasts, without using the actual measured power spectra, although where exactly the different contours intersect in parameter space is, of course, determined by the measurements.

As a point of comparison, a simple inverse variance weighting of the one-dimensional $H_0$ constraints from \planck, ACTPol, and SPTpol, ignoring the parameter correlations, yields $71.4\pm1.9$~km~s$^{-1}$~Mpc$^{-1}$. The reduction in the uncertainty in the full multidimensional fit illustrates this complementarity of the lower and higher multipoles of the EE spectrum for breaking parameter degeneracies.

While we do not yet have access to the SPT-3G EE likelihood, we note that the SPT-3G and SPTpol two-dimensional \lcdm\ parameter contours exhibit similar degeneracy directions \citep[Fig.~9 of][]{dutcher/etal:prep}. We therefore expect the combination of \planck\ and SPT-3G to produce similar results to the \planck+SPTpol or \planck+ACTPol results in Table~\ref{table:results}, including a lower value of $H_0$ and $n_s$ than reported for SPT-3G EE alone.

\begin{table}
  \centering
  \caption{Consistency of \lcdm\ parameters from different EE spectra, using test described in Section~\ref{sec:resultsconsistency}}
  \begin{tabular}{lcl}
\hline
\hline
EE Data Sets&Overall&Worst 1-D\\
\hline
\planck\ vs ACTPol&$1.0\sigma$&$0.7\sigma$ $(\Omega_bh^2)$\\
\planck\ vs SPTpol&$1.2\sigma$&$2.5\sigma$ $(A_se^{-2\tau})$\\
ACTPol vs SPTpol&$0.7\sigma$&$1.3\sigma$ $(A_se^{-2\tau})$\\
\hline
\planck\ vs ACTPol+SPTpol&$1.4\sigma$&$2.2\sigma$ $(A_se^{-2\tau})$\\
\planck+ACTPol vs SPTpol&$0.9\sigma$&$2.2\sigma$ $(A_se^{-2\tau})$\\
\planck+SPTpol vs ACTPol&$0.9\sigma$&$0.8\sigma$ $(\Omega_ch^2)$\\
\hline
\label{table:consistency}
\end{tabular}
\end{table}

\subsection{Consistency of different EE spectra within $\Lambda$CDM}
\label{sec:resultsconsistency}

In this subsection we quantify the consistency of the \planck, ACTPol, and SPTpol EE results within \lcdm. Provided the data sets are independent and the posterior parameter distributions are well-approximated as multivariate Gaussian\footnote{We discuss these assumptions in Sections~\ref{sec:resultsconsistencyindep} and \ref{sec:resultsconsistencygauss}.} we can perform a simple $\chi^2$ test for the consistency of the difference of the posterior means with zero \citep[e.g.,][]{addison/etal:2016,raveri/hu:2019},
\be
\chi^2_{12}=\sum_{ij}\left(\mu_{1,i}-\mu_{2,i}\right)\left(C_{11}+C_{22}\right)^{-1}_{ij}\left(\mu_{1,j}-\mu_{2,j}\right).
\ee
Here the data sets are labeled with subscripts 1 and 2, the Roman subscripts label parameters, and $\mu$ and $C$ are the mean and covariance estimated from the MCMC chains. This $\chi^2$ is converted to an equivalent Gaussian `$N\sigma$' by matching the probability-to-exceed from the $\chi^2$ distribution to that from a standard Gaussian distribution but considering only positive values (so $\chi^2=0$ corresponds to $0\sigma$ difference). These $N\sigma$ values are reported in Table~\ref{table:consistency}. Based on calculations using subsets of the full MCMC chains, the finite number of steps used to estimate $C$ sets an uncertainty floor of around $0.1\sigma$ in this test. Since the $\tau$ posteriors are driven by the common \texttt{lowE} likelihood we perform a comparison in five dimensions\footnote{Substituting $H_0$ for $\theta_{\rm MC}$ in these tests impacts the consistency results at $<0.1\sigma$.}, $\{\Omega_bh^2, \Omega_ch^2, \theta_{\rm MC}, A_se^{-2\tau}, n_s\}$, following, for example, \cite{aiola/etal:2020} and \cite{dutcher/etal:prep}.

In addition to testing the consistency between each pair out of \planck, ACTPol, and SPTpol, we tested the consistency between each experiment and the combination of the other two. These results are also shown in Table~\ref{table:consistency}. Overall we find no significant evidence for disagreement, with parameter differences no larger than $1.4\sigma$. This indicates that the preference for the lower $H_0$ value in the combined fit cannot be attributed to any significant tensions between the separate EE data sets.

The most notable single parameter difference is in $A_se^{-2\tau}$ for SPTpol, which falls $2.5\sigma$ low of \planck\ EE. The SPTpol polarization maps were calibrated by comparing the SPTpol and SPTpol$\times$\planck\ EE spectra on the SPTpol patch \citep[Section~7.3 of][]{henning/etal:2018}. This $2.5\sigma$ difference may therefore simply arise from an unfortunate statistical fluctuation on the SPTpol patch, but could also hint at some systematic issues arising when reanalyzing \planck\ data on small sky patches \citep[see also Section 13.3 of][]{choi/etal:2020}. Given the low statistical significance we do not attempt to investigate this issue further in this work. The SPT-3G data produced $A_se^{-2\tau}$ constraints in better agreement with \planck\ \citep[Fig.~9 of][]{dutcher/etal:prep}, although this is not surprising given the calibration was performed against the full-sky \planck\ spectra in that analysis. 

\subsubsection{Independence of EE spectra from different experiments}
\label{sec:resultsconsistencyindep}

In reality the different data sets are correlated with one another due to partial sky overlap. Due to a combination of current EE noise levels and the small size of the SPTpol patch we argue below that we can safely neglect these correlations (as done in previous studies). 

The covariance between the parameters from the \planck\ and ACTPol EE spectra was estimated by \cite{aiola/etal:2020} under the assumption that the E-modes measured by ACTPol are a subset of the modes accessible to the \planck\ analysis. \cite{aiola/etal:2020} found that ignoring the covariance would artificially tighten parameter constraints from a joint EE fit only at the per cent level (i.e., the uncertainties derived in a joint MCMC ignoring the covariance would be too tight by order per cent). This would fall below the uncertainty floor mentioned above for our consistency tests.

The SPTpol 500~deg$^2$ survey area is around nine times smaller than ACTPol's, covering only a few per cent of the area used in the \planck\ analysis, and thus we can also neglect the \planck-SPTpol EE covariance.

There is partial overlap between the SPTpol patch and the W5 field used in the ACTPol analysis below declination $-50^{\circ}$ (compare Fig.~1 of \citealt{henning/etal:2018} and Fig.~2 of \citealt{choi/etal:2020}). However, this overlap region is a small fraction of the total ($>4000$~deg$^2$) area used in the ACTPol analysis, and among the shallowest, with all the deep ACTPol fields lying at higher declination.

\subsubsection{Gaussianity of posterior distributions}
\label{sec:resultsconsistencygauss}

The ACTPol contours in Figure~\ref{fig:results} display clear non-Gaussian features, which arise primarily because the nuisance parameter $y_p$, which controls the polarization efficiency, is assigned a broad uniform prior on $[0.9,1.1]$. When fitting only to ACTPol EE data this opens up a large degeneracy with $A_se^{-2\tau}$ and, to some extent, other parameters. In joint fits with either TT or TE spectra, or another experiment, this degeneracy is broken and $y_p$ is tightly constrained, consistent with unity to within $1-2\%$ \citep[See Table~4 of][for results from joint fits to ACTPol and \wmap\ or \planck]{aiola/etal:2020}. For the consistency tests involving a comparison with ACTPol alone in Table~\ref{table:consistency} we therefore ran ACTPol chains fixing $y_p=1$, which produces approximately Gaussian posteriors for the \lcdm\ parameters (shown in the upper triangle of Fig.~\ref{fig:results}). This yields slightly more stringent consistency tests in the sense that the additional scatter corresponding to different $y_p$ values is not allowed.

Some small asymmetry of the one-dimensional posteriors is apparent for various parameters (not only for ACTPol) in Table~\ref{table:results}. Given the lack of evidence for tension between the data sets in Table~\ref{table:consistency} we have not pursued additional tests attempting to account for this.

\subsection{Comparison with \planck\ TT constraints and impact of gravitational lensing on the EE spectrum}
\label{sec:resultsplancktt}

Comparisons between parameter constraints from the CMB temperature and E-mode fluctuations are important for checking the performance of the \lcdm\ model and looking for hints of deviations. Since the \planck\ TT data is far more constraining than other TT results we compare the combined \planck+ACTPol+SPTpol EE parameters discussed above to the \planck\ TT parameters derived in conjunction with the \texttt{lowE} constraint on $\tau$ \citep{planck2016-l06}, rather than performing a joint TT analysis with ACT or SPT data.

Taking the EE and \planck\ TT constraints as independent, we find that the five-dimensional consistency test used in Table~\ref{table:consistency} yields TT-EE consistency at the $0.8\sigma$ level, with the largest single parameter difference being $1.4\sigma$ for $\Omega_bh^2$. The TT and EE spectra and parameters from the same sky area are partially correlated (evidenced by the non-zero TE spectrum). Over the multipole range accessible to \planck, however, the correlation coefficients between \lcdm\ parameters from TT and their EE counterparts have magnitude 0.1 or smaller, even for ideal noiseless data \citep[see Figs.~4 and 5 of][]{kable/addison/bennett:2020}. Given the significant noise levels in the \planck\ polarization data the TT and EE constraints are therefore well approximated as independent.

The \planck\ TT data has shown a persistent $>2\sigma$ preference for a larger-than-expected value of the phenomenological lensing amplitude parameter, $A_L$, when this is added as a free model parameter \citep[see][for most recent results]{planck2016-l06,efstathiou/gratton:2019}. This is associated with tensions between \lcdm\ parameters for different multipole ranges and a preference for a closed universe when using TT data alone \citep[e.g.,][]{planck2014-a13,addison/etal:2016,planck2016-LI,divalentino/melchiorri/silk:2020,efstathiou/gratton:2020,handley:2021}.

The \planck\ EE spectrum alone yields $A_L=1.32^{+0.24}_{-0.27}$. Combining the EE data sets produces a much tighter constraint, comparable in precision to TT, but shifted to lower values:
\begin{align}
\nonumber A_L&=0.89\pm0.10\\&\textrm{(\planck\ EE+ACTPol EE+SPTpol EE+\texttt{lowE})}\\
\nonumber A_L&=1.243\pm0.096\\\nonumber&\textrm{(\planck\ TT + \texttt{lowE})}.
\end{align}
These values differ at $2.5\sigma$, although the difference across the full parameter space (five \lcdm\ parameters, as in the earlier consistency tests, plus $A_L$) is $1.7\sigma$. The \lcdm\ parameters are in good agreement, differing only at $0.6\sigma$.

These results are qualitatively similar to those presented for the SPTpol TE+EE data by \cite{henning/etal:2018}, where the value of $A_L=0.81\pm0.14$ fell $2.9\sigma$ low of the \planck\ TT constraint. Marginalizing over $A_L$ produced good agreement for the \lcdm\ parameters from \planck\ TT and SPTpol, however, which were in mild tension for $A_L=1$.

\cite{aiola/etal:2020} reported $A_L=1.01\pm0.10$ for the ACTPol spectra (TT+TE+EE), and combining ACTPol and \wmap\ likewise resulted in a value centred around unity. \cite{story/etal:2013} reported $A_L=0.86^{+0.15}_{-0.13}$ from a joint fit to \wmap\ and the SPT TT spectrum. We find $A_L=1.18\pm0.15$ from a similar joint fit to \wmap\ and ACTPol TT.

Overall, current EE measurements are consistent with the \lcdm\ results from the \planck\ TT data, and do not show the same preference for $A_L>1$. The exact origin of the \planck\ TT $A_L$ behavior is unclear, and it may well have no connection to any underlying physics \citep[see also, e.g.,][]{couchot/etal:2017,efstathiou/gratton:2019}. Any modified cosmology models that do attempt to address the TT $A_L$ issue should also be tested against the EE data from \planck, ACTPol and SPTpol, however, given that \planck\ EE provides a far weaker $A_L$ constraint than the combination.

\subsection{Choice of prior on optical depth}
\label{sec:resultstauprior}

The results shown in this work adopt the \texttt{lowE} \planck\ likelihood to constrain $\tau$, based on a cross-correlation analysis of \planck\ High Frequency Instrument (HFI) 100 and 143~GHz data, using 30 and 353 GHz maps to clean the Galactic synchrotron and dust \citep[see Section~2.2 of][for more details]{planck2016-l05}. In recent years, a number of other studies have constrained $\tau$ using different combinations of \planck\ and \wmap\ data, including alternative processing and mapmaking for the \planck\ polarization \citep[e.g.,][]{weiland/etal:2018,planck2020-LVII,beyondplanck1,natale/etal:2020}. These analyses give a spread in mean values of $\tau$ from 0.05 to 0.07, and $1\sigma$ uncertainties from 0.006 to 0.2. The \texttt{lowE} likelihood gives $\tau=0.0506\pm0.0086$ \citep{planck2016-l06}. We also ran MCMC chains instead using a Gaussian prior $\tau=0.065\pm0.015$, matching the choice adopted in the ACTPol analysis by \cite{aiola/etal:2020}. We found that for the EE results shown in Table~\ref{table:results} the impact of this choice is very small, with shifts in posterior means at the $0.1\sigma$ level, and changes in 68\% interval bounds only at the few per cent level.

\section{Conclusions}
\label{sec:conclusions}

We have examined the \lcdm\ parameter constraints from separate and joint fits to EE power spectra from the \planck, ACTPol, and SPTpol surveys, motivated by the recent observation that the EE spectra from each experiment separately produce higher values of $H_0$ than, for example, \planck\ TT, in good agreement with the Cepheid-SNe SH0ES ladder \citep{dutcher/etal:prep}.

A joint fit to \planck, ACTPol and SPTpol EE spectra yields $H_0=68.7\pm1.3$~km~s$^{-1}$~Mpc$^{-1}$. This value is $2.4\sigma$ lower than the distance ladder, and also lower than the result from any of the data sets separately. This behavior arises from different degeneracy directions across the full \lcdm\ parameter space, particularly for $\Omega_bh^2$ and $n_s$. We found, however, that the EE spectra from the different experiments produce consistent \lcdm\ parameters, with differences across the five-dimensional parameter space (excluding $\tau$) at the $1.4\sigma$ level or lower (Table~\ref{table:consistency}). In other words, the shift to a lower $H_0$ is not because the data sets are incompatible. There is a $2.5\sigma$ tension in values of the power spectrum amplitude, $A_se^{-2\tau}$, inferred from \planck\ EE and SPTpol EE, which may hint at some calibration issues.

The \lcdm\ parameters from the joint EE fit are consistent with the \lcdm\ parameters from the \planck\ TT data within $0.8\sigma$. We found that the EE data precisely constrain the phenomenological lensing amplitude parameter $A_L$, preferring a value consistent with unity, $A_L=0.89\pm0.10$. Like earlier analyses of the ACTPol and SPTpol data we do not reproduce the preference for $A_L>1$ seen in the \planck\ TT spectrum.

Based on our results, the preference for higher $H_0$ values from the separate EE measurements seems more likely to be due to chance fluctuations than the first hint of systematic differences in preferred parameters from, for example, the TT data. Such differences could still exist, of course, and will no doubt be the subject of future work with upcoming data. New theoretical models that impact the TT and EE spectra differently may yet be promising for resolving the Hubble disagreement. Given the consistency between the current TT and EE data within \lcdm, however, it seems unlikely that such models would be favored over \lcdm\ at a statistically significant level by CMB data.

\section*{Acknowledgments}

I would like to thank Chuck Bennett for many valuable discussions, as well as comments on this work. I'm also grateful to Janet Weiland, Gary Hinshaw, and Mark Halpern for helpful discussions and suggestions, and to Erminia Calabrese and Jason Hennings for help with the ACTPol and SPTpol likelihood codes.

This work was supported in part by NASA ROSES grants NNX17AF34G and 80NSSC19K0526. This work was based on observations  obtained with \Planck\ (http://www.esa.int/Planck), an ESA science mission with instruments and contributions directly funded by ESA Member States, NASA, and Canada. I acknowledge the use of the Legacy Archive for Microwave Background Data Analysis (LAMBDA), part of the High Energy Astrophysics Science Archive Center(HEASARC). HEASARC/LAMBDA is a service of the Astrophysics Science Division at the NASA Goddard Space Flight Center. This research project was conducted using computational resources at the Maryland Advanced Research Computing Center (MARCC).

\software{GetDist v1.1.2 \citep{lewis:2019}, NumPy v1.19.5 \citep{vanderwalt/colbert/varoquaux:2011}, Matplotlib v3.3.3 \citep{hunter:2007}, SciPy v1.6.0 \citep{virtanen/etal:2020}}


\end{document}